\documentclass[letter]{aa}  

\usepackage{graphicx}
\usepackage{xcolor}
\usepackage{siunitx}

\usepackage{chemformula}
\usepackage{natbib}
\usepackage{multirow}
\usepackage{colortbl}
\usepackage{lscape}
\usepackage{placeins}

\newcommand{\Msun}{\ensuremath{\mathrm{M}_\odot}\xspace}

\newcommand{\aMLT}{\ensuremath{\alpha_\mathrm{MLT}}\xspace}
\newcommand{\FeH}{\ensuremath{[\ch{Fe}/\ch{H}]}\xspace}

\newcommand{\dYdZ}{\ensuremath{\Delta Y/\Delta Z}\xspace}
\newcommand{\Xc}{\ensuremath{X_\mathrm{c}}\xspace}
\newcommand{\tcoll}{\ensuremath{t_\mathrm{coll}}\xspace}

\makeatletter
\renewcommand*\aa@pageof{, page \thepage{} of \pageref*{LastPage}}
\makeatother

\usepackage{hyperref}
\hypersetup{colorlinks=true, allcolors=blue}

\begin{document}

   \title{Predictions of gravity mode pulsations of collisional blue straggler stars in globular clusters}

   \subtitle{}

   \author{L. Briganti\inst{\ref{inst1}, \ref{inst2}}\fnmsep\thanks{Corresponding author: \href{mailto:lorenzo.briganti2@unibo.it}{lorenzo.briganti2@unibo.it}}
    \and W. E. van Rossem\inst{\ref{inst1}}
    \and A. Miglio\inst{\ref{inst1}, \ref{inst2}}
    \and A. Bragaglia\inst{\ref{inst2}}
    \and M. Matteuzzi\inst{\ref{inst1}, \ref{inst2}}
    }

   \institute{Department of Physics \& Astronomy "Augusto Righi", University of Bologna, via Gobetti 93/2, 40129 Bologna, Italy\label{inst1}
         \and
            INAF-Astrophysics and Space Science Observatory of Bologna, via Gobetti 93/3, 40129 Bologna, Italy\label{inst2} 
             }

   \date{Received ; accepted }
 
  \abstract 
   {Blue straggler stars (BSSs) are exotic objects, which, being the results of processes such as mass transfer, mergers, or collisions, are considered key objects in the study of their host clusters' dynamics. While many studies on astrometric, spectroscopic, and photometric properties of BSSs in clusters have been conducted, there are few works in the literature regarding their pulsations and internal structure, which can indeed retain traces of their origin. In this work we computed and analysed a grid of collisional BSSs at low metallicity ($Z = 0.01\; Z_\odot$), finding that collision products present a peculiar chemical stratification that leads to periodicities in the period-spacing pattern of high-order gravity modes. These seismic fingerprints provide a unique opportunity to   constrain the formation pathways of BSSs in globular clusters.}


   \maketitle
   \nolinenumbers

\section{Introduction}
\label{sec: Introduction}
Blue straggler stars (BSSs) are core-hydrogen burning stars which lie, in the colour-magnitude diagram (CMD), in a bluer and brighter extension of the main sequence (MS), thus appearing younger compared to the rest of the stars in their host clusters. Originally found in globular clusters (GCs; see \citealt{sandage1953}), they have also been observed in open clusters (OCs; see, e.g. \citealt{rain2021}), in dwarf spheroidal galaxies (see, e.g. \citealt{momany2015}), and among field stars (see, e.g. \citealt{preston2015} for a review). \par

Until now, three main formation channels have been suggested. Accretion by mass transfer in  binary systems \citep{mccrea1964, ivanova2015}, collisions or mergers in triple systems \citep{andronov2006, perets&fabricky2009}, and direct collisions \citep{hills&day1976, leonard1989}. While the first and second formation channels are dominant in less dense environments, such as OCs \citep{rain2024} or low-density GCs \citep{sollima2008, ferraro2025}, where multiple stellar systems have higher production and survival probabilities, the third one plays a major role in more crowded environments, such as GC and OC cores \citep{wang&ryu2024}. However, these three mechanisms can also coexist, and a combination of them is needed to explain, for example, the existence of double sequences of BSSs in post core-collapse GCs (see, e.g. \citealt{cadelano2022}). \par

In GCs, properties such as radial distribution (see, e.g. \citealt{ferraro2020}) and rotational velocities (see \citealt{billi2024} and references therein) have been extensively studied, revealing that BSSs are  powerful indicators of the dynamical stage of these clusters. On the other hand, there is still the need to investigate the internal structure of BSSs, not only to better constrain their formation pathways, but also to more easily discover these objects among field stars or inside clusters' MSs, if they did not acquire sufficient mass to exceed the MS turn-off (MSTO); that is the case for blue lurkers \citep[BLs;][]{leiner2019, leiner2025, nine2023}. In this framework, a crucial role could be played by asteroseismology \citep[see, e.g.][]{aerts2021} and its ability to probe stellar interiors through  the study of oscillation frequencies. In particular, gravity (g) modes represent a precious tool to investigate the interiors of MS stars, from chemical stratification \citep{miglio2008} to rotation frequency (see, e.g. \citealt{vanreeth2016} and \citealt{ouazzani2017}) and angular-momentum transport \citep{aerts2019}. This means that g modes can be used as probes of the chemical structure of stars and to determine whether collision products retain traces of their dynamical past in their interiors. Pulsating BSSs in the instability regions of $\gamma$~Doradus ($\gamma$~Dor; see \citealt{kaye1999} and \citealt{handler1999}), $\delta$ Scuti, or SX Phoenicis stars have been studied in OCs by \cite{guzik2023} and \cite{vernekar2023}, as well as in the stellar system $\omega$ Centauri by \cite{daszynska-dasziewicz2025}.
Furthermore, \cite{hatta2021} reports a first example of an asteroseismic modelling of a likely field BSS exhibiting $\gamma$~Dor pulsations, showing the potential of combining asteroseismology and parametric models to describe the interiors of objects with a modified structure, such as BSSs. \par

In this work, following what was done by \cite{henneco2025} for massive stars, we modelled low-metallicity collisional products using the Make Me a Star (\verb|MMAS|; \citealt{lombardi2002}) software and compared their internal structure and oscillations to single-stellar-evolution (SSE) objects of equivalent masses, focusing on high-order g modes, typical of $\gamma$~Dors. In Sect.~\ref{sec: Methods}, we introduce our grids of SSE and BSS models and briefly describe the properties of g modes and their influence from chemical stratifications. In Sect.~\ref{sec: Results_and_discussion}, we present and discuss our results. Finally, we draw our conclusions in Sect.~\ref{sec: Conclusions}.

\section{Methods}
\label{sec: Methods}
We computed SSE models and merged them to produce a grid of collisional BSSs. We then computed the g-mode frequencies for both sets of models and compared the evolutionary tracks, internal profiles, and g-mode pulsations of SSE and BSS models with the same mass.

\subsection{Model grid}
\label{subsec: Grid}

We used the stellar evolution code Modules for Experiments in Stellar Astrophysics (\verb|MESA|, version 24.08.1; see \citealt{jermyn2023} and references therein) and the Python package \verb|wsssss|\footnote{\url{https://github.com/waltervrossem/wsssss}} to compute a grid of SSE non-rotating models with masses between $0.45\; \Msun$ and $1.8\; \Msun$ (with steps of $0.01\; \Msun$) and fixed $Z = 0.01\; Z_\odot$ ($\sim\FeH = -2\; \si{dex}$, a metallicity that can be representative of stars in GCs). All the specifications adopted for stellar evolution are reported in Appendix~\ref{app_subsec: stellar evolution}. \par

We then computed a grid of collisional BSSs using \verb|MMAS| (see Appendix~\ref{app_subsec: MMAS} for more details), as done in \cite{glebbeek2008} and \cite{glebbeek&pols2008} for OCs. We used two MS profiles from the
SSE grid as input profiles for \verb|MMAS|  and considered head-on collisions, to avoid the effects of rotation (see Appendix~\ref{appendix: off-axis collisions}). We set the MS for SSE objects as the phase where the central $\ch{H}$-mass-fraction, $\Xc$, was between $0.99 X_\mathrm{c,init}$ and $10^{-4}$. After obtaining the profile of the collision product, we performed a smoothing operation applying a Savitzky-Golay filter \citep{savitzky&golay1964} by means of the Python package \verb|SciPy| \citep{virtanen2020}. We then used the smoothed profile as an input for \verb|MESA| and evolved the BSS with the same input physics used for SSE models. \par

The BSS grid was constructed such that the SSE model with mass equal to the sum of the parent masses lay within the typical effective-temperature range of $\gamma$~Dor stars (\citealt{dupret2004}; Fellay et al., in prep.). For each selected mass, $M$, we varied the primary-star mass, $M_1$, between $M/2$ and $M-0.45\; \Msun$ (with $0.45\; \Msun$ being the minimum mass in our SSE grid) and set the secondary-star mass to $M_2 = M - M_1$. In this way, for each possible BSS, we obtained different combinations of parent masses and, consequently, different mass ratios $q=M_2/M_1$. The final masses obtained for the BSSs are slightly lower (by $\sim 6\%$) than $M_1+M_2$ because of the mass lost during the collision (see Appendix~\ref{app_subsec: MMAS}). The time of the collision (i.e. the parent stars' age at the moment of the collision), $t_\mathrm{coll}$, was set to ensure that the BSSs were in the MS phase at $t=12.86\; \si{Gyr}$, which is the median age of GCs with $-2.1\; \si{dex} \leq \FeH \leq -1.9\; \si{dex}$ in \citealt{kruijssen2019}. In the end, our BSS grid was composed of 49 models with masses going from $0.901\; \Msun$ to $1.105\; \Msun$ (compatible with mass estimates from \citealt{billi2024}) and $q$ ranging from 0.625 to 1. \par

\subsection{Asteroseismology}
\label{subsec: Asteroseismology}
As previously mentioned, since BSSs in GCs populate the $\gamma$~Dor instability strip, in this paper we focused on exploring $\gamma$~Dor-type modes, i.e. high-order g modes. Gravity modes are pulsation modes whose main restoring force is buoyancy. They propagate in radiative zones and decay exponentially in convective regions. Their characteristic frequency is the Brunt-Väisälä (BV) frequency \citep{vaisala1925, brunt1927}; i.e. the oscillation frequency of a local mass element displaced from its equilibrium position:
\begin{equation}
    N^2 = g\biggl(\frac{1}{\Gamma_1}\frac{\mathrm{d}\ln P}{\mathrm{d}r}-\frac{\mathrm{d}\ln\rho}{\mathrm{d}r}\biggr),
    \label{eq: Brunt_Vaisala_def}
\end{equation}
where $g$ is the gravitational acceleration, $\Gamma_1 = (\partial\ln P/\partial\ln \rho)_\mathrm{ad}$ the first adiabatic exponent, $r$ the distance from the stellar centre, and $P$ and $\rho$ pressure and density, respectively. \par
In the asymptotic regime (i.e. when $n \gg \ell$, with $n$ being the radial order and $\ell$ the angular degree), assuming chemically homogeneous, non-rotating, and non-magnetic stars \citep{tassoul1980}, g-modes with the same $\ell$ and consecutive values of $n$ are equally spaced in period and the period spacing, $\Delta P = P_{n-1} - P_n$, is proportional to the buoyancy travel time, or total buoyancy radius, defined as
\begin{equation}
    \Pi_0 = 2\pi^2 \biggl(\int_{r_1}^{r_2} \frac{N(r')}{r'} \mathrm{d}r' \biggr)^{-1},
    \label{eq: total_buoyancy_radius}
\end{equation}
where $r_1$ and $r_2$ correspond, respectively, to the inner and outer turning points of the g-mode propagation cavity (the region where $N^2$ is positive). If the integration in Eq.~\ref{eq: total_buoyancy_radius} is performed between $r_1$ and a generic position $r$ between $r_1$ and $r_2$, we obtain the expression for the local buoyancy radius, $\Pi_r$. The normalised buoyancy radius, $\Pi_0/\Pi_r$, serves as a radial coordinate in the g-mode propagation cavity. \par

Given that the BV frequency depends on the chemical-composition gradient, $\nabla_\mu$, and that the asymptotic period spacing depends on the buoyancy radius (see, e.g. \citealt{aerts2021}), the period spacing is sensitive to deviations from the assumption of homogeneous chemical profile. In particular, as shown by \cite{kawaler1995} for white dwarfs and by \cite{miglio2008} for MS stars with convective cores, a steepening in the chemical profile produces a sharp peak in $\nabla_\mu$ and, consequently, in $N^2$. This makes the g-mode period-spacing pattern (PSP) a tool for investigating chemical anomalies in BSSs compared to the SSE objects of the same mass. \par

We then computed, both for SSE and BSS models, adiabatic oscillation frequencies for dipole ($\ell = 1$) modes using the stellar oscillation code \verb|GYRE| \citep{townsend&teitler2013}, version 8.0. We looked for oscillation modes in a period interval compatible with observations from \cite{li2020}. Finally, we compared BSS and SSE models of the same mass, looking at the evolution of the PSPs, hydrogen composition profiles and BV frequency profiles for different values of the central hydrogen mass-fraction.

\section{Results and discussion}
\label{sec: Results_and_discussion}

\begin{figure}
    \centering
    \includegraphics[width=0.95\columnwidth]{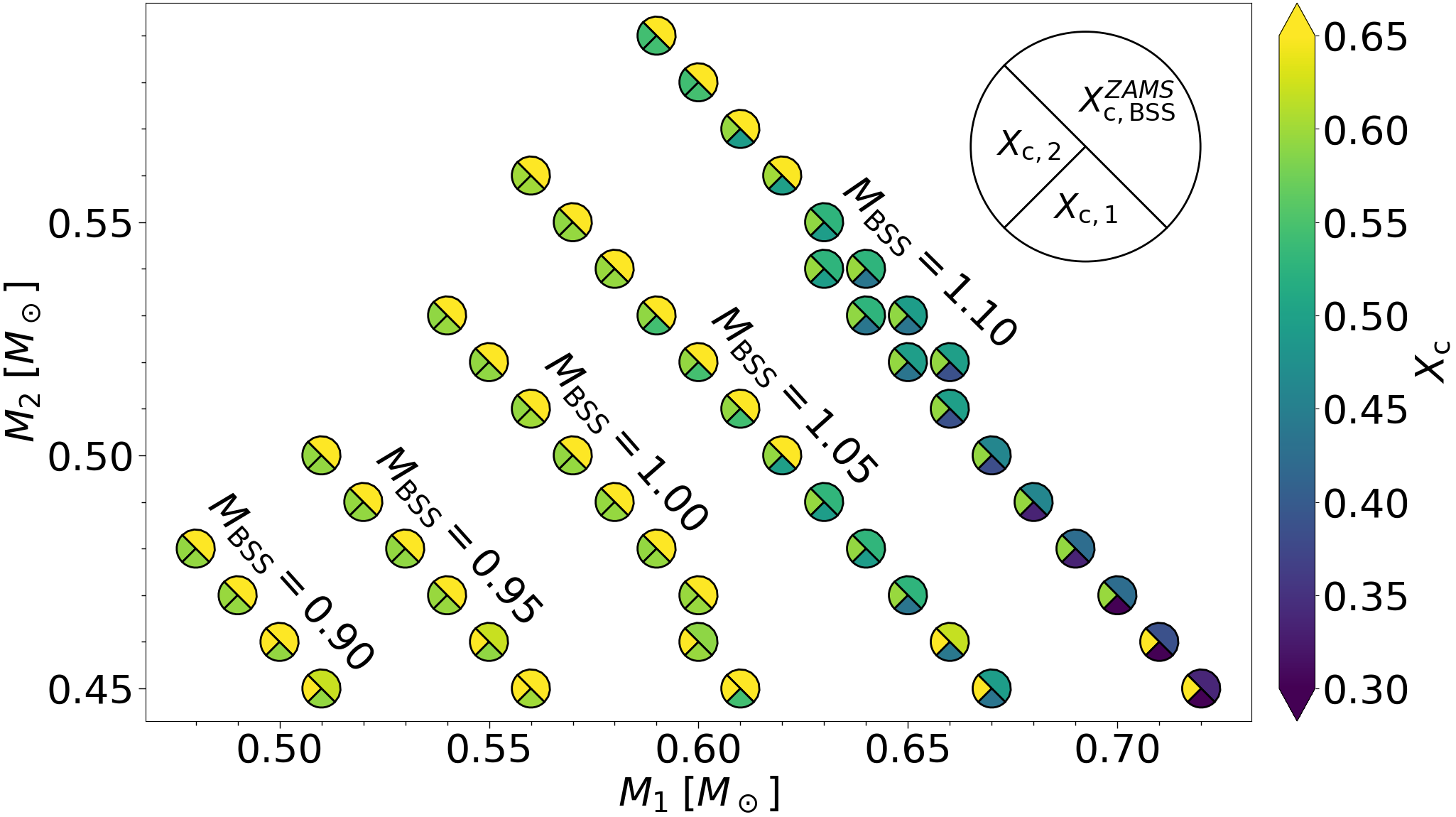}
    \caption{Representation of BSS grid. The x-axis and y-axis report all the different values of $M_1$ and $M_2$, respectively. Each point is divided into three sectors, indicating the values of $\Xc$ for the primary (bottom sector) and the secondary star (left sector) at the time of the collision, and the value of $\Xc$ for the BSS at the zero age MS (ZAMS; top right sector). The colour-coding for $\Xc$ is reported on the left. The colour-bar ranges from 0.3 to 0.65. Values out of this range are coloured as the closest extreme on the colour scale. The BSSs' masses are reported inside the plot, rounded to the closest multiple of $0.01$.}
    \label{fig: BSS_grid}
\end{figure}

\begin{figure}
    \centering
    \includegraphics[width=0.95\columnwidth]{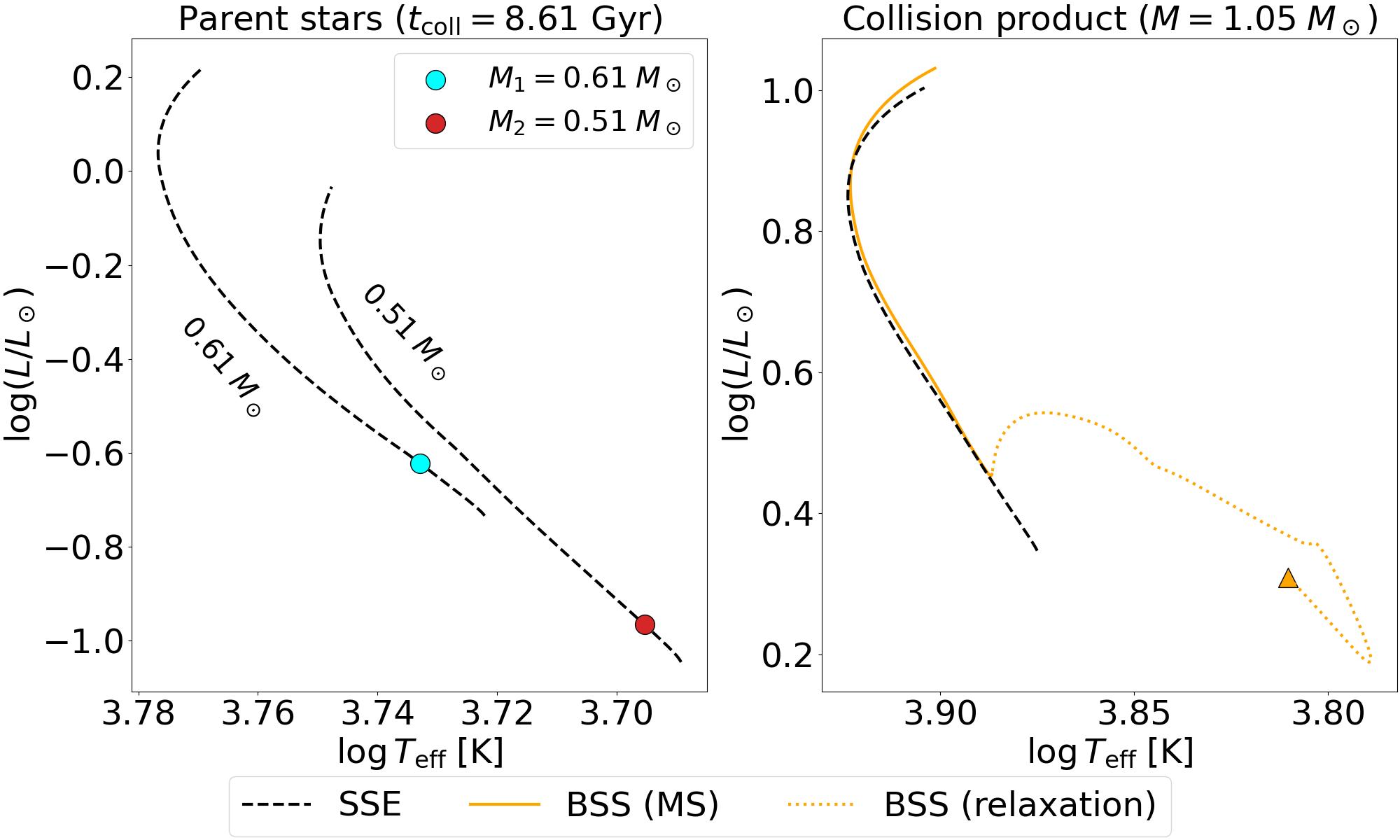}
    \caption{HRDs of collision between an $M_1 = 0.61\; \Msun$ star and a $M_2 = 0.51\; \Msun$ star. Left panel: Evolutionary tracks of parent stars (dashed black lines), with the position of the parent stars at the moment of the collision (cyan for the primary, red for the secondary). Right panel: Evolutionary tracks of BSS (orange; dotted lines = relaxation, solid lines = MS) and of the equivalent-mass SSE star (black dashed line). The triangle marks where the relaxation phase starts.}
    \label{fig: tracks_0.61_0.51}
\end{figure}

\begin{figure*}
   \centering
   \includegraphics[width=17cm]{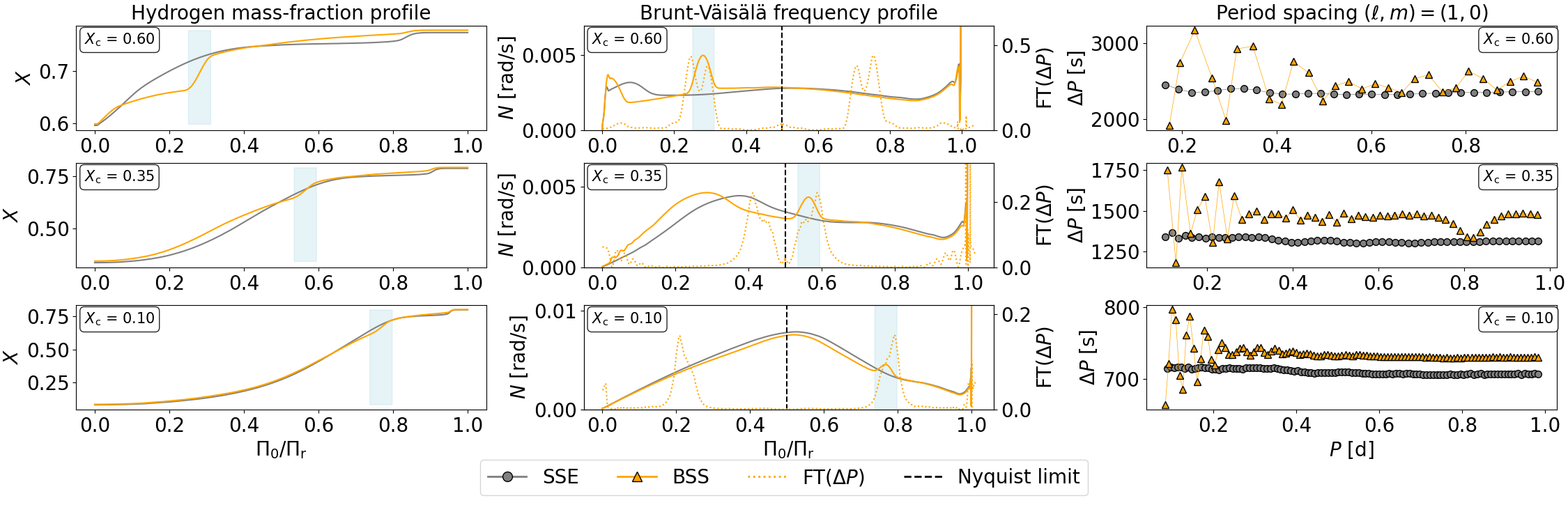} 
   \caption{Comparison between a $1.05\; \Msun$ BSS ($M_1 = 0.61\; \Msun$, $M_2 = 0.51\; \Msun$; orange) and its SSE counterpart (grey) for different values of $\Xc$ (0.60, 0.35, and 0.10; from top row to bottom row). Left column: Evolution of $X$ profile in g-mode cavity as a function of normalised buoyancy radius, $\Pi_0/\Pi_r$. Centre column: Evolution of $N$ as function of $\Pi_0/\Pi_r$ (solid lines). The dotted orange line represents the FT of the BSS's PSP, with the Nyquist limit ($\Pi_0/\Pi_r = 0.5$) indicated by the dashed black line. Right column: Evolution of PSPs for dipole modes (note that the scales on the period-axis are not the same for all the panels). In the left and centre panels, a light blue shaded region marks the position of the glitch, which corresponds to the steepening of the $X$ profile and to the additional peak in the $N$ profile.}
    \label{fig: PSP_glitch_0.61_0.51}
\end{figure*} 

We first describe the general properties of our grid, and then discuss the comparison between a BSS and an SSE object with equivalent mass for a specific case. Similar results for other cases are reported in Appendices~\ref{appendix: HRDs} and~\ref{appendix_PSPs}. \par

The main effect of varying $q$ at fixed BSS mass ($M_\mathrm{BSS}$) and $\tcoll$ is that it changes the parent stars' evolutionary stage. In particular, for lower values of $q,$ the primary star is more massive and, consequently, more evolved and poorer in hydrogen with respect to primary stars in  cases of higher mass ratios. Analogously, the secondary star is less massive and, consequently, richer in hydrogen. This has effects on the initial chemical composition of the BSS. Specifically, in our grid we observe that a decrease in $q$ produces BSSs with lower values of $\Xc$ (see Fig.~\ref{fig: BSS_grid}). \par

Before evolving along the MS, BSSs undergo a relaxation phase during which the structure contracts, heats up, and develops a convective core. Similarly to what we did for SSE stars (see Sect.~\ref{subsec: Grid}), we set the start of the MS to the time when $\Xc$ was equal to 99\% of its maximum value reached during the relaxation phase. In all cases, the relaxation phase lasts for less than $100\; \si{Myr}$. Examples of the evolution in the Hertzsprung-Russell diagram (HRD) of some BSSs in our grid are provided in Fig.~\ref{fig: tracks_0.61_0.51} and in Appendix~\ref{appendix: HRDs}. Notably, BSSs' tracks are almost indistinguishable from those of SSE models of the same mass. \par

Despite these nearly identical surface properties, the internal chemical stratification of BSSs differs significantly, and this contrast leaves a clear seismic imprint. Indeed, as shown in Fig.~\ref{fig: PSP_glitch_0.61_0.51} (and the same holds for the entire grid; other examples are reported in Appendix~\ref{appendix_PSPs}), the $\ch{H}$-mass-fraction of the BSS, differently from its SSE counterpart, presents an additional steepening within the radiative zone. This steepening, as discussed in \cite{miglio2008}, produces sharp peaks in the chemical gradient and in the BV profile, leading to a periodicity in the PSP. The presence of an additional peak in the BV frequency profile of collision products has also been shown by \cite{henneco2025}, in the case of massive stars. We verified that the periodicity in the PSP is associated with the peak in the BV profile using both the Fourier transform (FT) of the PSP (see, e.g. \citealt{guo2025}) and the comparison between the local wavelength of g modes, $\lambda_\mathrm{g}$, and the BV frequency scale height, $H_N$ (see, e.g. \citealt{cunha2020, hatta2023}, and \citealt{matteuzzi2025}). During the evolution along the MS, the steepening in $X$ and the additional peak in $N$ reduce in intensity, leaving signatures on progressively shorter g-mode oscillation periods, and move through the external part of the g cavity. \par 

Finally, we varied the periastron radius of the orbit, $r_\mathrm{P}$, for the specific case analysed in this section. As expected from the formulation adopted in \verb|MMAS|, we found that off-axis collisions produce BSSs with higher values of the specific angular momentum, as discussed in Appendix~\ref{appendix: off-axis collisions}, suggesting a framework to explore possible angular momentum distributions and internal rotation, which can be inferred using asteroseismic constraints.

\section{Conclusions}
\label{sec: Conclusions}

We present the first seismic analysis of low-metallicity collisional BSS models, comparing them with SSE stars of equal mass. Our results show that collision products follow very similar evolutionary tracks with respect to their SSE counterparts. Most importantly, we demonstrate the potential of g-mode pulsations to probe BSS interiors. We find clear seismic signatures of collisions in the theoretical spectra: the altered chemical stratification of BSSs produces an additional slope change in the chemical composition profile, an additional peak in the BV frequency, and periodic modulations in the pulsation spectrum. These features provide a direct probe of their internal structure and open the way to constraining BSS formation channels with asteroseismology. Extending the analysis to higher metallicities, different impact parameters, and to the study of internal rotation could prove essential to finding further evidence of their distinct structure and history. Equally important will be systematic comparisons with existing data from {\it Kepler} \citep{borucki2010, gilliland2010} and with observations from future or candidate missions such as PLATO \citep{rauer2025} and HAYDN \citep{miglio2021}.

\begin{acknowledgements}
We warmly thank the referee for providing valuable comments, which pushed us to undertake further discussions, resulting in a more coherent version of the manuscript. LB acknowledges financial support from MUR (Ministero dell'Università e della Ricerca) and NextGenerationEU throughout the PNRR ex D.M. 118/2023. WEvR, AM, and MM acknowledge support from the ERC Consolidator Grant funding scheme (project ASTEROCHRONOMETRY, https://www.asterochronometry.eu, G.A. n. 772293). LB acknowledges Conny Aerts, Gaël Buldgen, Mario Cadelano, Marc-Antoine Dupret, Zhao Guo, Yoshiki Hatta, Alex Kemp, and Gang Li for the useful discussions that helped to improve the work.
\end{acknowledgements}

\bibliographystyle{aa}
\bibliography{References.bib}

\appendix

\section{Stellar models}
\label{appendix: stellar models}
In this appendix we describe in detail the prescriptions we adopted in the stellar evolution code \verb|MESA| and how the software \verb|MMAS| works. 

\subsection{Stellar evolution}
\label{app_subsec: stellar evolution}
As already mentioned in Sec.~\ref{subsec: Grid}, we used \verb|MESA|, version 24.08.1, to compute the SSE grid and evolve the collision products obtained from \verb|MMAS|. We adopted the solar metal mixture from \cite{asplund2009} and obtained from the solar calibration $Y_{\odot, \mathrm{init}} = 0.268$ and $Z_{\odot, \mathrm{init}} = 0.016$ as initial abundances for the Sun and a mixing length parameter $\aMLT = 1.97$, according to the formulation in \cite{cox&giuli1968}. Assuming a linear enrichment law and setting $Y_\mathrm{P} = 0.2485$ \citep{komatsu2011}, we obtained an enrichment ratio $\dYdZ = 1.218$. \par

We included atomic diffusion, as well as a turbulent mixing prescription described in \citet[see Eq. 1]{richer2000}. Furthermore, we adopted an exponential convective overshooting \citep{herwig2000}, with the temperature gradient adjustment from \cite{michielsen2023} and setting $D_\mathrm{ov}^\mathrm{min} = 10^{-2}\; \si{cm^2/s}$ as the minimum diffusion coefficient, as suggested in \cite{buchele2025}. The overshooting efficiencies have been set to $f_\mathrm{ov, core} = 0.018$ for the convective core and $f_\mathrm{ov, env} = 0.025$ for the convective envelope in both cases. Finally, for the atmosphere boundary conditions, we used the $T(\tau)$ relation presented in \cite{ball2021} and based on \cite{trampedach2014} The inlists and \verb|run_star_extras.f| are available online\footnote{\url{https://doi.org/10.5281/zenodo.17735309}}. \par

\subsection{Make Me A Star}
\label{app_subsec: MMAS}

\begin{figure}[h!]
   \centering
   \includegraphics[width=\columnwidth]{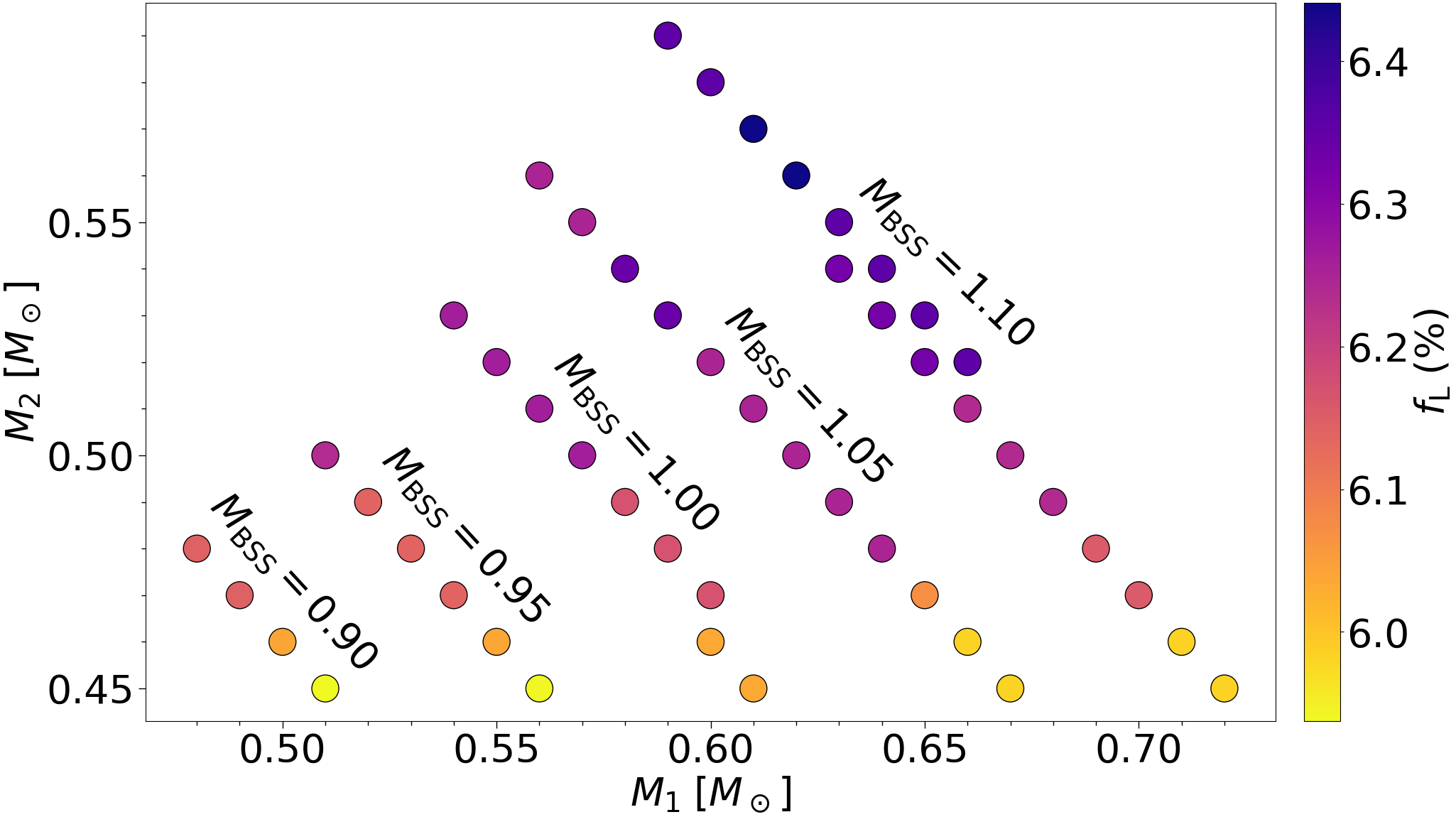} 
   \caption{Representation of our BSSs' grid with points coloured by the percentage of mass lost during the collision.}
    \label{fig: Grid_mass_loss}
\end{figure}

The starting profiles of the BSSs have been computed using \verb|MMAS| \citep{lombardi2002}, which is based on approximation of results from smoothed particle hydrodynamics. This code takes as inputs the profiles of the parent stars at the time of the collision and the periastron separation of the initial parabolic orbits ($r_\mathrm{P}$) and performs as a main operation an entropy sorting, i.e. a process that takes into account that fluid particles with lower specific entropy sink to the stellar centre. This is done by ordering each layer from the parent stars such that the entropic variable $A$ increases monotonically from the centre to the surface. The entropic variable $A$ is defined as $A = P/\rho^{\Gamma_1}$ (see Eq. 1 from \citealt{glebbeek2008} for its dependence on specific entropy). Since entropy, hence $A$, is constant in convective regions (except for the outer layers of the envelope, in which the thermal gradient is super-adiabatic), elements coming from convective zones of the same parent stars will also be contiguous in the collision product. In addition,
\verb|MMAS| takes into account the mass-loss and shock heating due to the collision, with formulations based on results from smoothed particle hydrodynamics simulations. In particular, the mass-loss is expressed as a fraction of the sum of the parent stars' masses, $M_1$ and $M_2$:
\begin{equation}
    \begin{split}
        f_L &= \frac{(M_1+M_2)-M_\mathrm{BSS}}{M_1+M_2} = \\
    &= c_1\frac{q}{(1+q)^2}\frac{R_{1,0.86}+R_{2,0.86}}{R_{1,0.5}+R_{2,0.5}+c_2r_\mathrm{P}},
    \end{split}
    \label{eq: mass_loss}
\end{equation}
where $c_1 = 0.157$ and $c_2 = 1.8$ are fixed parameters, and $R_{i,0.5}$ and $R_{i,0.86}$ are the radii of the $i$-th parent star containing $50\%$ and $86\%$ of its mass, respectively.
In our grid the mass-loss is around $\sim 6\%$, with a slight increase for higher values of $q$ (see Fig.~\ref{fig: Grid_mass_loss}). Finally, shock heating produces an increase of the entropic variable $A$, dependent on a function of $r_\mathrm{P}$ and on the pressure $P$ of each layer (see Sec. 2.2 from \citealt{lombardi2002}).

\FloatBarrier

\clearpage

\section{Off-axis collisions}
\label{appendix: off-axis collisions}
As mentioned in Sec.~\ref{subsec: Grid}, we built our BSSs grid considering only head-on collisions, to not consider the effects of rotation. We discuss in this appendix the consequences of changing the impact parameter of the collision. \par

\begin{figure}[h!]
   \centering
   \includegraphics[width=\columnwidth]{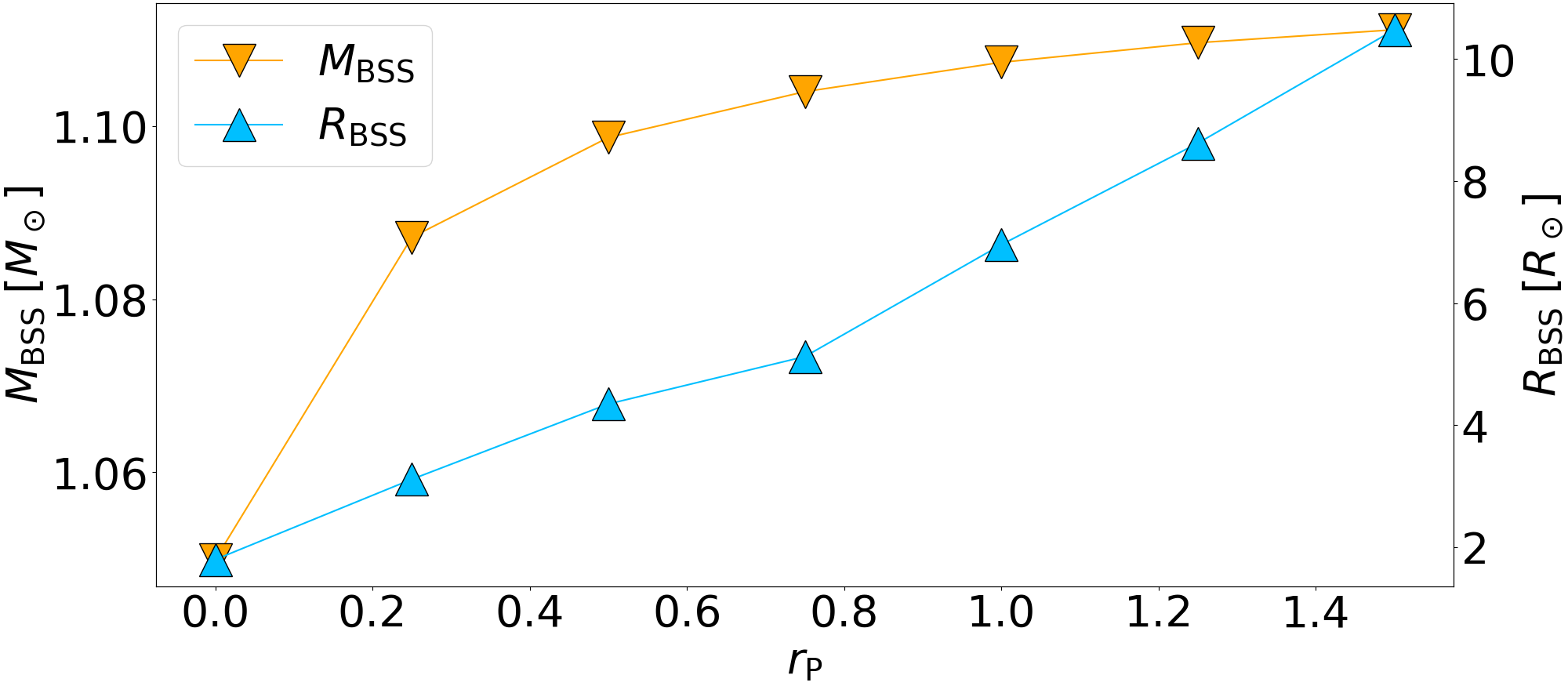} 
   \caption{Mass (orange) and radius (light blue) of a collision product ($M_1 = 0.61\; \Msun$, $M_2 = 0.51\; \Msun$) as functions of the periastron radius, $r_\mathrm{P}$.}
    \label{fig: M_R_vs_rp}
\end{figure}

In \verb|MMAS| it is possible to set the value of the periastron radius of the initial parabolic orbit, $r_\mathrm{P}$, in units of the sum of the radii of the parent stars, respectively $R_1$ and $R_2$. The possible values range from 0 (head-on collision) to 2. The first dependence on this parameter is related to the mass lost in the collision. As described by Eq.~\ref{eq: mass_loss}, the mass-loss is inversely proportional to $r_\mathrm{P}$. This means that \verb|MMAS| produces more massive BSSs for off-axis collisions (i.e., for higher values of $r_\mathrm{P}$), when the parent stars are fixed (see Fig.~\ref{fig: M_R_vs_rp}). We also observed that for off-axis collisions the BSS has a larger radius (see Fig.~\ref{fig: M_R_vs_rp}). \par

\FloatBarrier

\begin{figure}[h!]
   \centering
   \includegraphics[width=\columnwidth]{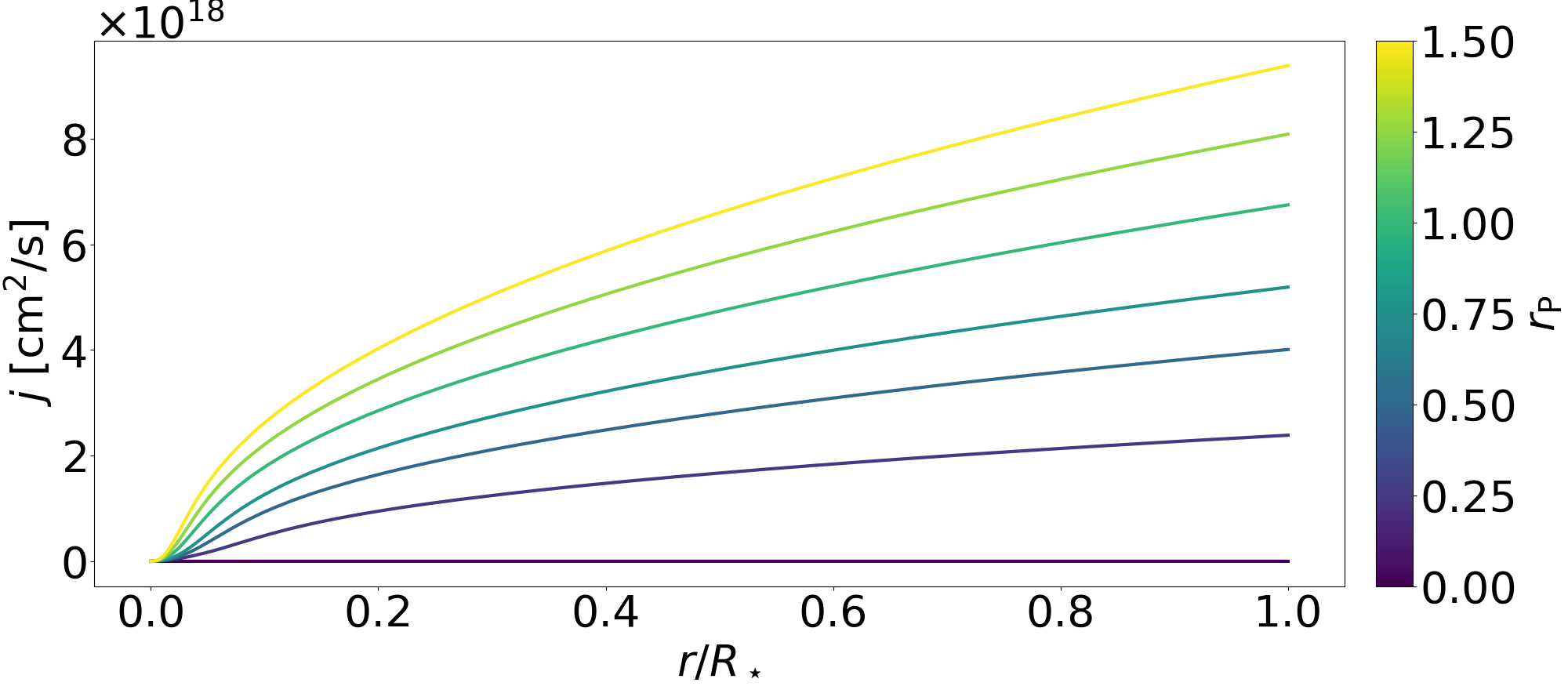} 
   \caption{Profiles of the specific angular momentum, $j$, of a collision product with $M_1 = 0.61\; \Msun$ and $M_2 = 0.51\; \Msun$, coloured by the value of the periastron radius, $r_\mathrm{P}$.}
    \label{fig: j_profile}
\end{figure}

The choice of $r_\mathrm{P}$ has consequences also on the rotation of the BSS. Specifically, the total angular momentum of the collision, $J_\mathrm{tot}$, which is the sum of the BSS' angular momentum, $J_\mathrm{BSS}$, and a component due to mass-loss, increases with increasing $r_\mathrm{P}$:
\begin{equation}
    \begin{split}
        J_\mathrm{tot} &= J_\mathrm{BSS} + c_9 f_\mathrm{L}J_\mathrm{tot} = \\
        &= M_1 M_2 [2Gr_\mathrm{P}/(M_1+M_2)]^{1/2},
    \end{split}
    \label{eq: angular momentum}
\end{equation}
where $c_9 = 2$ is a fixed parameter, $f_\mathrm{L}$ is defined by Eq.~\ref{eq: mass_loss}, $G$ is the gravitational constant, and $M_1$ and $M_2$ are the parent stars' masses.
From Eq.~\ref{eq: angular momentum} we observe that, while for head-on collisions we obtain non-rotating products, in the case of off-axis collisions, for increasing values of $r_\mathrm{P}$, the specific angular momentum, $j = J_\mathrm{BSS}/m$ (with $m$ being the mass coordinate), increases, producing BSSs with higher rotational velocities (see Fig.~\ref{fig: j_profile}).

\FloatBarrier

\section{Hertzsprung-Russell diagrams}
\label{appendix: HRDs}
We report in this appendix additional examples of HRDs of collisions for different values of the mass-ratio, $q$, and of the final mass, $M_\mathrm{BSS}$.

\begin{figure}[h!]
   \centering
   \includegraphics[width=\columnwidth]{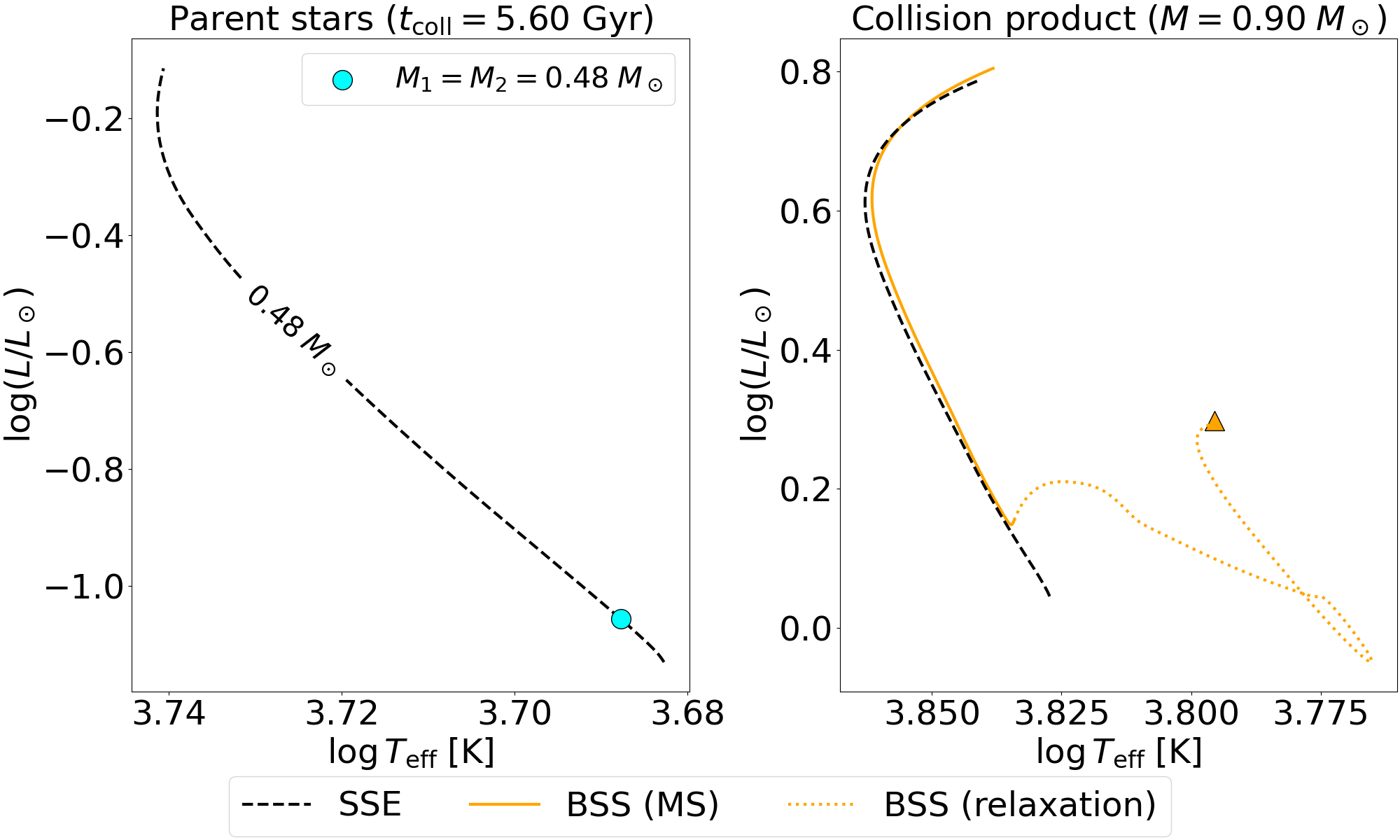} 
   \caption{Same as Fig.~\ref{fig: tracks_0.61_0.51} but for a $0.90\; \Msun$ BSS ($M_1 = M_2 = 0.48\; \Msun$). Given that the primary and the secondary stars occupy the same position in the HRD, we omitted the red point in the left panel.}
    \label{fig: tracks_0.48_0.48}
\end{figure} 

\begin{figure}[h!]
   \centering
   \includegraphics[width=\columnwidth]{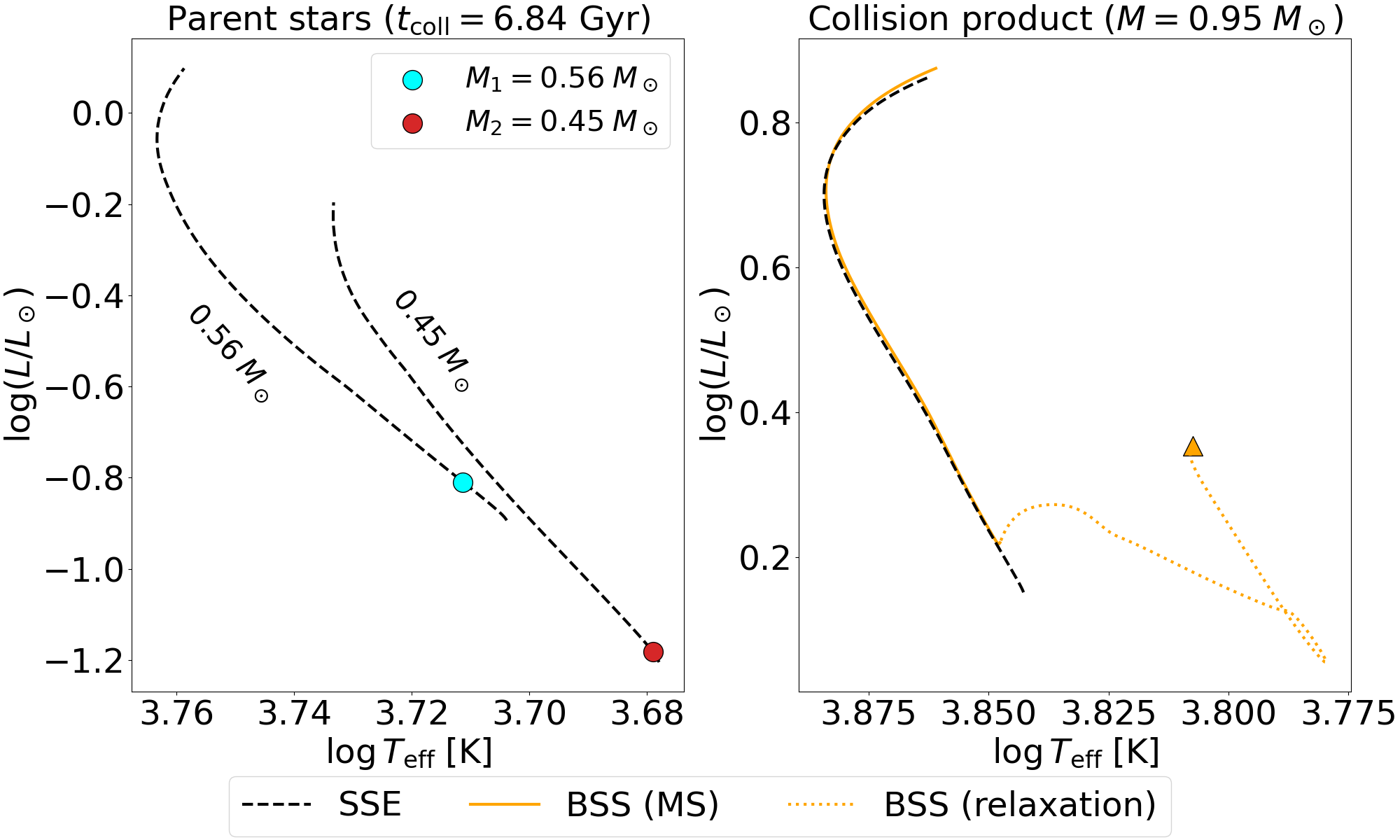} 
   \caption{Same as Fig.~\ref{fig: tracks_0.61_0.51} but for a $0.95\; \Msun$ BSS ($M_1 = 0.56\; \Msun$, $M_2 = 0.45\; \Msun$).}
    \label{fig: tracks_0.56_0.45}
\end{figure} 

\begin{figure}[h!]
   \centering
   \includegraphics[width=\columnwidth]{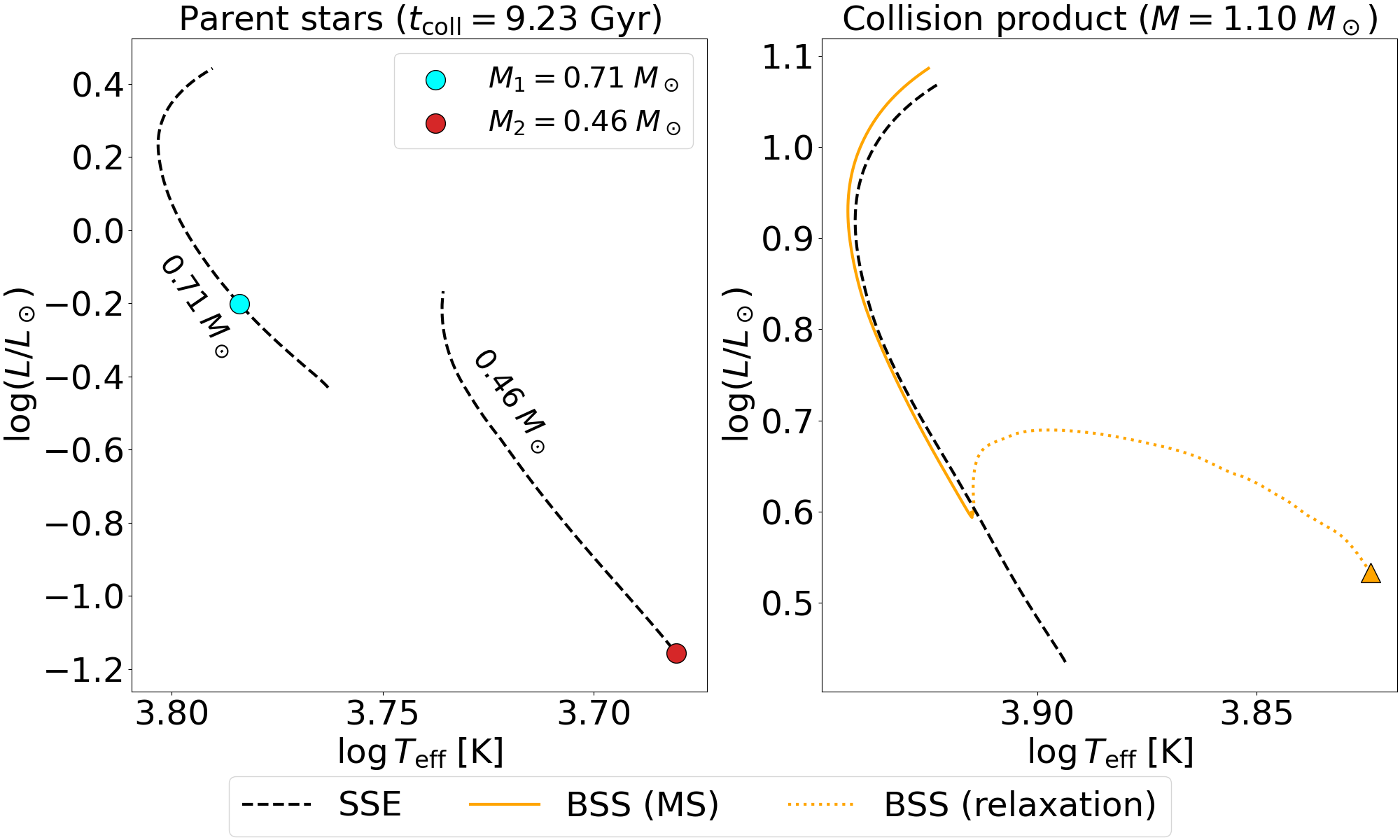} 
   \caption{Same as Fig.~\ref{fig: tracks_0.61_0.51} but for a $1.10\; \Msun$ BSS ($M_1 = 0.71\; \Msun$, $M_2 = 0.46\; \Msun$).}
    \label{fig: tracks_0.71_0.46}
\end{figure} 
\FloatBarrier

\onecolumn
{
\section{Internal profiles and period spacing patterns}
\label{appendix_PSPs}
We provide in this appendix the comparison between BSSs and their SSE counterparts on the chemical and BV frequency profiles in the g-mode cavity, and on the PSP, for different values of the mass ratio in the collision.  

\begin{figure*}[h!]
   \centering
   \includegraphics[width=17cm]{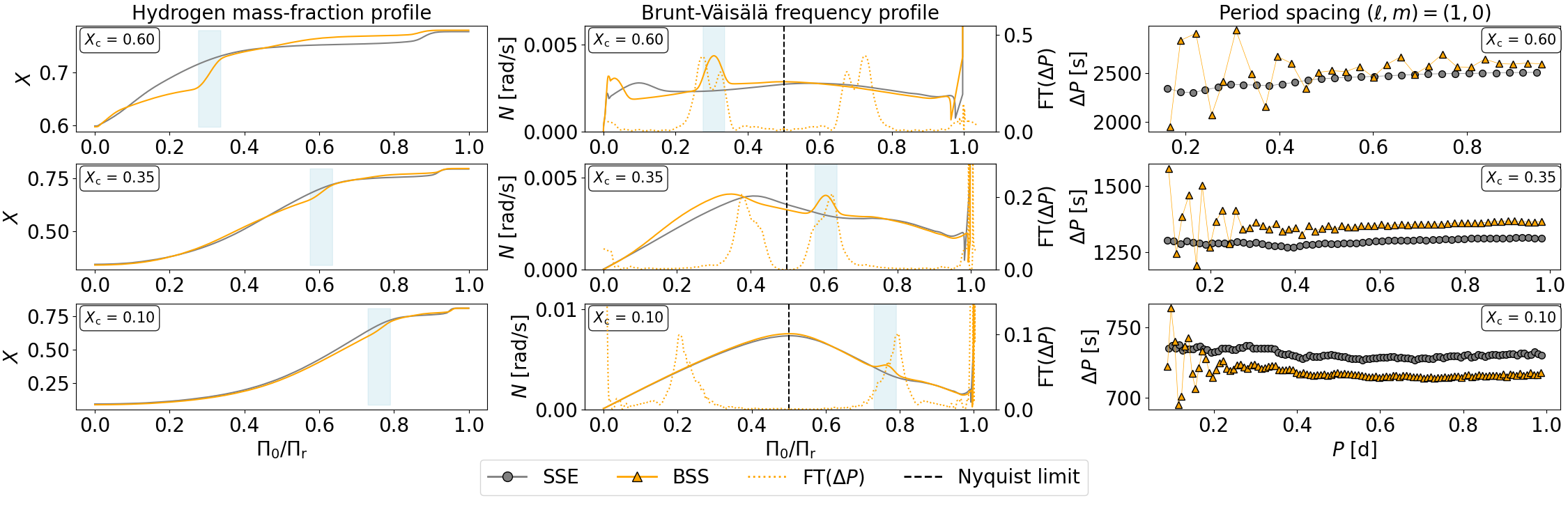} 
   \caption{Same as Fig~\ref{fig: PSP_glitch_0.61_0.51} but for a $0.90\; \Msun$ BSS ($M_1 = M_2 = 0.48\; \Msun$).}
    \label{fig: PSP_0.48_0.48}
\end{figure*}

\begin{figure*}[h!]
   \centering
   \includegraphics[width=17cm]{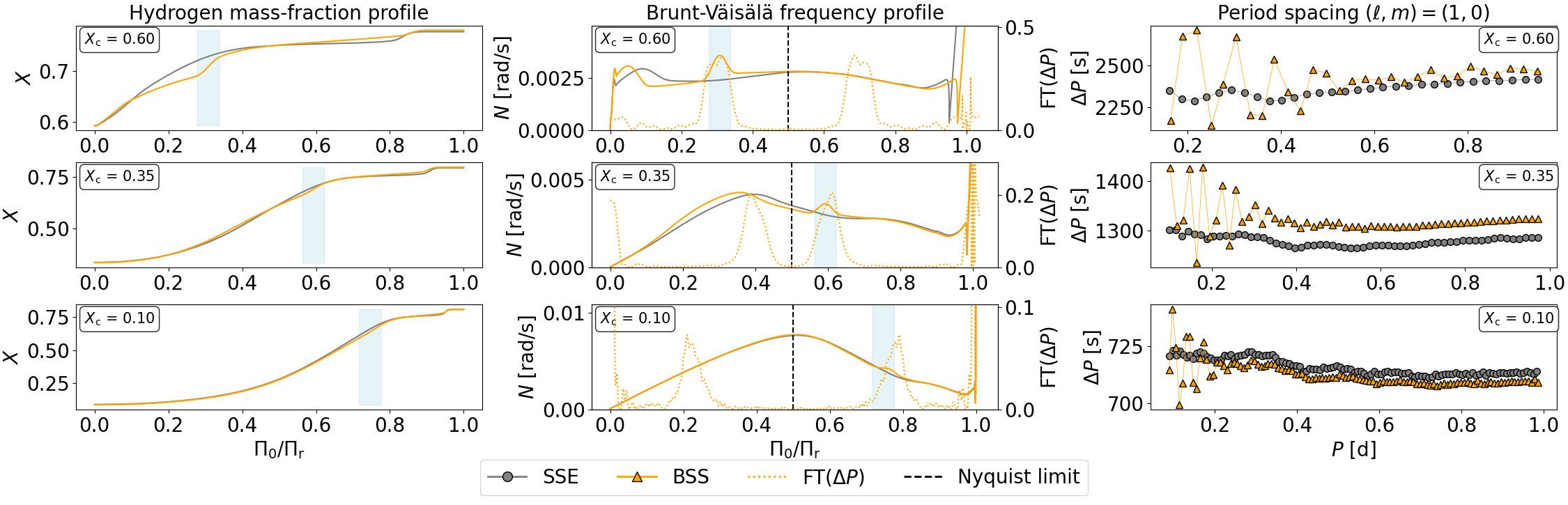} 
   \caption{Same as Fig~\ref{fig: PSP_glitch_0.61_0.51} but for a $0.95\; \Msun$ BSS ($M_1 = 0.56\; \Msun$, $M_2 = 0.45\; \Msun$).}
    \label{fig: PSP_0.56_0.45}
\end{figure*}

\begin{figure*}[h!]
   \centering
   \includegraphics[width=17cm]{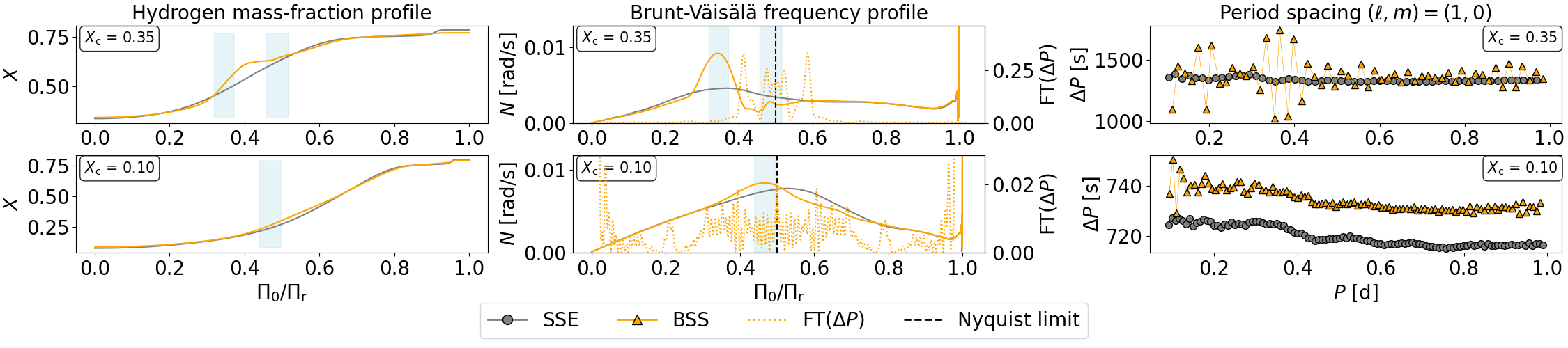} 
   \caption{Same as Fig~\ref{fig: PSP_glitch_0.61_0.51} but for a $1.10\; \Msun$ BSS ($M_1 = 0.71\; \Msun$, $M_2 = 0.46\; \Msun$) and only for $\Xc = 0.35, 0.10$. In the first row we highlighted two glitches (two steepenings in the $X$ profile, two peaks in the BV profile), which produce beatings in the PSP.}
    \label{fig: PSP_0.71_0.46}
\end{figure*}
\FloatBarrier
}

\label{LastPage}
\end{document}